\documentclass{article}
\usepackage{spconf,amsmath,graphicx}
\usepackage{siunitx}
\usepackage{bm,bbm}
\usepackage{amsfonts,amssymb,amsmath}
\usepackage{booktabs}       %
\usepackage{pifont}%
\usepackage{makecell}
\usepackage{multirow}
\usepackage{balance} %
\usepackage{url}
\usepackage{hyperref}

\def\@fnsymbol#1{\ensuremath{\ifcase#1\or *\or \dagger\or \ddagger\or
   \mathsection\or \mathparagraph\or \|\or **\or \dagger\dagger
   \or \ddagger\ddagger \else\@ctrerr\fi}}
\newcommand{\ssymbol}[1]{^{\@fnsymbol{#1}}}

\title{Adapting Speech Separation to Real-World Meetings\\Using Mixture Invariant Training}
\name{Aswin Sivaraman$^{1,2,*}$\thanks{$^*$Work done during an internship at Google.}, Scott Wisdom$^1$, Hakan Erdogan$^1$, John R.\ Hershey$^1$}
\address{
$^1$Google Research\qquad 
$^2$Indiana University\\ 
}
\begin{document}
\ninept
\maketitle
\begin{abstract}
The recently-proposed mixture invariant training (MixIT) is an unsupervised method for training single-channel sound separation models in the sense that it does not require ground-truth isolated reference sources. In this paper, we investigate using MixIT to adapt a separation model on real far-field overlapping reverberant and noisy speech data from the AMI Corpus. The models are tested on real AMI recordings containing overlapping speech, and are evaluated subjectively by human listeners. To objectively evaluate our models, we also devise a synthetic AMI test set. For human evaluations on real recordings, we also propose a modification of the standard MUSHRA protocol to handle imperfect reference signals, which we call MUSHIRA. Holding network architectures constant, we find that a fine-tuned semi-supervised model yields the largest SI-SNR improvement, PESQ scores, and human listening ratings across synthetic and real datasets, outperforming unadapted generalist models trained on orders of magnitude more data. Our results show that unsupervised learning through MixIT enables model adaptation on real-world unlabeled spontaneous speech recordings.
\end{abstract}
\begin{keywords}
source separation, unsupervised learning, mixture invariant training, real-world audio processing
\end{keywords}
\section{Introduction}
\label{sec:intro}

Extracting estimates of clean speech in the presence of interference is a long-standing research problem in signal processing.  This task is referred to as \textit{speech enhancement} when the interference is non-speech, or \textit{speech separation} when the interference can include speech.  While there has been tremendous progress in recent years with speaker-independent speech separation, made possible using deep learning algorithms \cite{LuoY2019convtasnet,Chen2020dual}, an important problem remains unsolved regarding the mismatch between training and test domains.

This mismatch problem stems in part from reliance on \textit{supervised training} of speech separation models \cite{LuoY2019convtasnet,Chen2020dual,HuangP2015ieeeacmaslp, hershey2016deep, yu2017permutation, wang2018supervised}.  
Supervised training requires access to clean reference signals, which are synthetically combined to form input mixtures.
In supervised training, clean reference signals are required as training targets. It is generally infeasible to obtain recordings of real mixtures along with clean references, due to cross talk between the microphones.  Thus it is necessary for supervised training to mix together clean signals into synthetic mixtures, for which the clean signals serve as references.  

Synthetic mixing exacerbates the problem of mismatch, because the generated mixtures have to be engineered by hand to match the distribution of the target domain on many dimensions, including speaker characteristics, speaker position and motion, speech activity patterns, noise types, noise patterns, and acoustic reverberation \cite{KinoshitaK2013REVERB, BarkerJ2015CHIME}. Currently this problem has been approached by creating data with enough variety that, it is hoped, some subset will match the target domain \cite{maciejewski2018building,manilow2019cutting}.   

Alternative training approaches directly address this mismatch by using matched noisy data from the test domain. One prior work proposed test-time adaptation for speech enhancement systems---personalizing the model by optimizing its parameters towards a particular speaker using only noisy speech recordings \cite{SivaramanA2021interspeech}. However, this approach relies on a large labeled non-speech dataset, which may not be obtainable within the target domain. 

Other recent works propose training exclusively on unlabeled real-world recordings, which may provide a better match to the target domain's characteristics.
One work is a generative model using a variational autoencoder \cite{neri2021unsupervised}, although this model was only validated on small domains such as mixtures of MNIST images and spectrograms of a few musical instrument mixtures. For more general unsupervised separation, the recently proposed \emph{mixture invariant training} (MixIT) \cite{wisdom2020mixit} enables discriminative training on raw mixture audio without labels or ground-truth reference signals.

One caveat to MixIT is that it involves using \emph{mixtures of mixtures} (MoMs) as inputs, which may potentially create a form of mismatch with the target domain, where the input is a single mixture. In contrast, MoMs have more active sources on average than single mixtures, and perhaps some inconsistency in the acoustics between two different recordings.  However, it is an empirical question whether the mismatch introduced by MixIT is as detrimental as the forms of mismatch that MixIT alleviates.  One way to mitigate both risks is to jointly train on supervised synthetic data, which may better approximate the target domain in terms of the number of sources and consistency of the reverberation.

Previous experiments showed that MixIT performed well at adaptation to reverberation \cite{wisdom2020mixit}.  However, these experiments were conducted using synthetic data as the target domain, and were compared with supervised training data that was strongly mismatched in terms of both source activity as well as reverberation.  Thus the benefit of such adaptation for real data remains to be verified.

In this paper, we explore the effectiveness of a neural network-based speech separation system in a real meeting room domain. In particular, we experiment with the AMI Corpus dataset \cite{CarlettaJ2006AMI}, where no matching supervised training data exists.  To this end, we train our system using either: (1) supervised training with synthetic reverberant data, or (2) unsupervised MixIT training using AMI data (i.e.\ matched domain), or (3) a \textit{semi-supervised} combination of the two.  Lastly, we also investigate the benefits of pretraining the model using MixIT on AudioSet \cite{AudioSet}, a very large open-domain dataset, prior to the above configurations or in isolation.

Evaluation of our models on real-world data presents a challenge: objective metrics cannot be used due to the lack of reference signals. To address this, we perform human listening tests using the real AMI Corpus test set data. To handle the lack of perfect reference signals for the real data, we propose an extension of the MUSHRA (multiple stimuli with hidden reference and anchors) \cite{itu2014mushra} protocol that we call MUSHIRA (multiple stimuli with hidden {\it imperfect} reference and anchors), where headset recordings containing some cross-talk are used as an imperfect reference. In order to measure objective metrics, we also construct a synthetic AMI mixture dataset, which utilizes the synchronized headset and distant microphone recordings in addition to word boundary annotations. To create this proxy dataset, we use a linear time-invariant filter to project audio from headset microphones to distant microphones, and create mixtures of these pseudo-references.  Our evaluations confirm that MixIT is helpful for adaptation, with the best results produced by a combination of supervised and unsupervised training.

\section{Training Methods}
\label{sec:litreview}
One approach to supervised source separation when sources are of the same or ambiguous class is to use permutation-invariant training (PIT) \cite{hershey2016deep, yu2017permutation}. Given a mixture $x$ with reference sources ${\bf s}\in\mathbb{R}^{M\times T}$ and separated sources $f_\theta(x)=\hat{\bf s}\in\mathbb{R}^{M\times T}$, the PIT objective is
\begin{equation}
\begin{aligned}
\mathcal{L}_\mathrm{PIT}\left({\bf s}, \hat{\bf s}
\right) = \min_{  {\bf P}} \, & \sum_{m=1}^{M}\mathcal{L}\left({s}_m, [{\bf P} \hat{\bf s}]_m\right),
\end{aligned}
\label{eq:pit}
\end{equation}
where ${\bf P}$ is an $M\times M$ permutation matrix and $\mathcal{L}$ is a signal-level loss function.

To train a source separation model on real-world data which lacks reference sources requires an unsupervised learning algorithm.  The recently proposed mixture-invariant training (MixIT) \cite{wisdom2020mixit} accomplishes this using a form of self-supervision. In this approach, the model inputs are mixtures of mixtures (MoMs), which are the sum of 2 reference mixtures, (i.e., ${\bar{x}} = {x}_1 + {x}_2$ where $x_{n} \in \mathbb{R}^{T}$). Given reference mixtures and separated sources $\hat{\bf s}=f_\theta(\bar{x})$, the MixIT loss estimates a mixing matrix $\mathbf{A} \in \mathbb{B}^{2 \times M}$:
\begin{equation}
\mathcal{L}_{\text{MixIT}}
\left(\left\{x_{n}\right\}, \hat{\mathbf{s}}\right)=\min _{\mathbf{A} \in \mathbb{B}^{2\times M}} \sum_{n=1}^2 \mathcal{L}\left(x_{n},[\mathbf{A} \hat{\mathbf{s}}]_{n}\right)
\end{equation}
\noindent where $\mathbb{B}^{2 \times M}$ is the set of $2 \times M$ binary matrices where each column sums to 1 (i.e., the set of matrices which assign each separated source $\hat{s}_{m}$ to one of the reference mixtures $x_{n}$), and $\mathcal{L}$ is a signal-level loss function between reference mixtures and their estimates.  Training is thus discriminative with respect to the individual mixtures, and the individual source estimates emerge as latent variables. 

In this paper, for both PIT and MixIT, we use negative thresholded SNR as the signal-level loss function:
\begin{equation}
\mathcal{L}(y, \hat{y})=-10 \log _{10} \frac{\|y\|^{2}}{\|y-\hat{y}\|^{2}+\tau\|y\|^{2}}
\end{equation}
where $\tau=10^{-\mathrm{SNR_{\max}} / 10}$ acts as a soft threshold that clamps the loss at $\mathrm{SNR}_{\max }$. We empirically select $\mathrm{SNR}_{\max }=30 \mathrm{~dB}$.

\section{Experiment Setup}
\label{sec:experiments}

Our speech separation model is the ``improved time-domain convolutional neural network'' (TDCN++) \cite{kavalerov2019universal}, which estimates a fixed number of masks for a particular basis representation of the input mixture. This input representation is multiplied by the masks, and the result is inverted back to a set of time-domain separated waveforms. For the basis representation, we either use a learned basis \cite{LuoY2019convtasnet} with \SI{2.5}{\ms} window and \SI{1.25}{\ms} hop, or a short-time Fourier transform (STFT) with \SI{32}{\ms} window and \SI{8}{\ms} hop. As our experiment pertains to domain adaptation through both supervised and unsupervised training, we hold this model architecture constant and vary training schemes.
All models were trained with the Adam optimizer \cite{kingma2014adam} with batch size 128 and learning rate $10^{-3}$ on 32 Google Cloud TPU v3 cores, with early stopping on the validation loss.

\begin{figure}
    \centering
    \includegraphics[width=\linewidth,trim={1.5cm 0 0 0},clip]{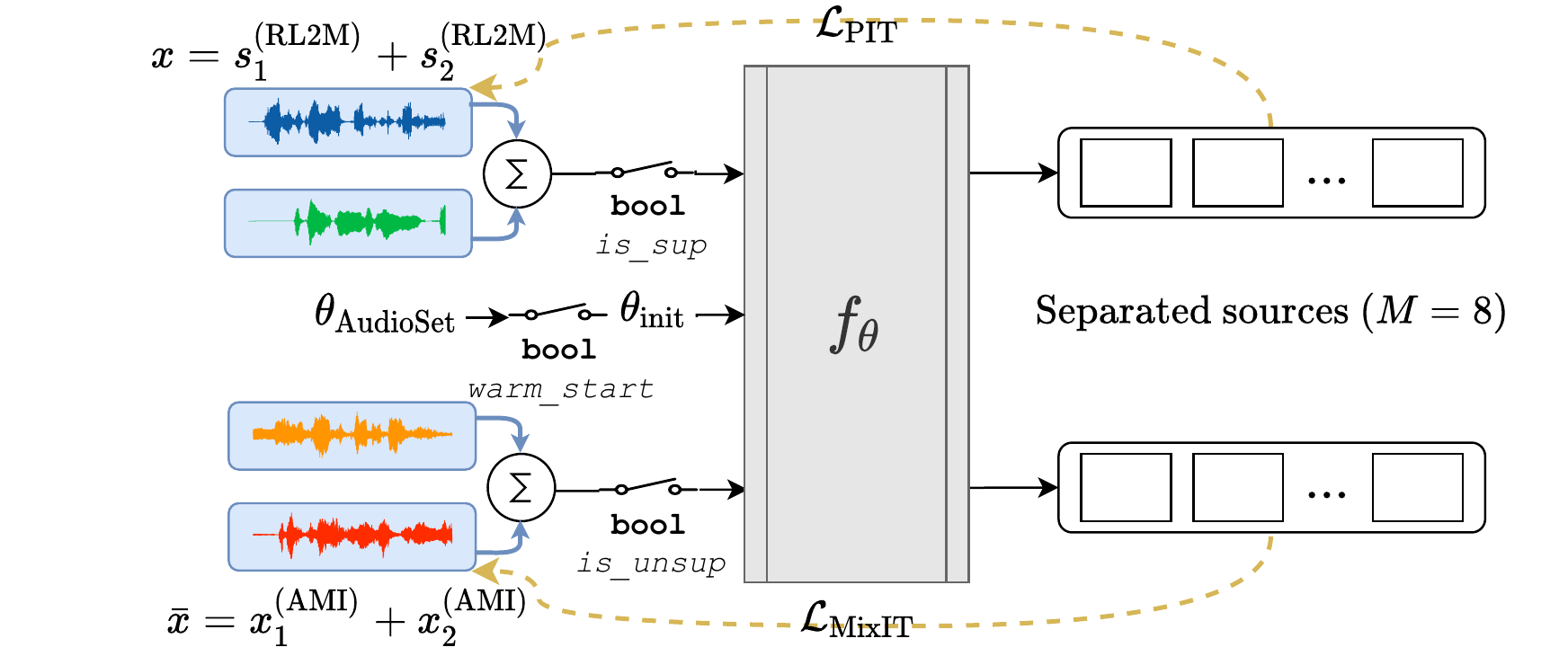}
    \caption{Training configuration diagram. Our experiments train a separation model ($f_\theta$) through combinations of supervised PIT with reverberant Libri2Mix (RL2M), unsupervised MixIT with AMI, and warm-start (parameter initialization) from a model pretrained using unsupervised MixIT on AudioSet.}
    \label{fig:diagram}
    \vspace{-10pt}
\end{figure}

Our training configurations are illustrated in Figure \ref{fig:diagram}. For supervised data, we use anechoic and reverberant versions of Libri2Mix \cite{cosentino2020librimix,wang2021sequential}. The anechoic version is the official clean two-speaker mixtures, and the reverberant version RLibri2Mix \cite{wisdom2020mixit} uses synthetic impulse responses using a simulator described in previous work \cite{wang2021sequential}.

For unsupervised data, raw \SI{20}{\second} clips of AMI Corpus distant microphone audio are used. During training, random \SI{10}{\second} clips are sampled from these raw \SI{20}{\second} clips to increase diversity of training examples. When creating MoMs from this raw AMI Corpus data, we only combine mixtures from the same location (Edinburgh, TNO, or Idiap). If this consistency between locations is not used, we found that the separation model works poorly in practice, since it exploits the provided cue during training.
As in recent work \cite{wisdom2021sparse}, we found that pretraining the separation model on AudioSet \cite{AudioSet} was very helpful. For all experiments, we used the same checkpoints trained for about 2.7M steps using MixIT on \SI{10}{\second} clips of AudioSet.

In terms of training dataset durations, supervised (reverberant) Libri2Mix is \SI{212}{\hour}, unsupervised matched AMI is \SI{34}{\hour} (Edinburgh), \SI{23}{\hour} (TNO), and \SI{14}{\hour} (Idiap), and unsupervised open-domain AudioSet is \SI{5800}{\hour}.
Note that the amount of AMI data is one order of magnitude less than the amount of supervised data, and two orders of magnitude less than the amount of AudioSet data.

\section{Evaluation}
\label{sec:datasets}

To estimate the objective performance of our methods, we use synthetic mixtures of AMI data,
as well as real AMI recordings. We also propose an extension of the MUSHRA listening test to handle imperfect reference signals.

\subsection{Synthetic Overlapping AMI}
\label{ssec:synthetic_ami}

To enable performance measurement for our methods using objective  measures on data that is exactly matched to AMI as much as possible, we constructed a synthetic evaluation set with isolated sources from the AMI recordings. As a first step, we used the provided transcript annotations to segment utterances based on word boundaries, which yielded \SI{17}{\hour} of isolated non-overlapping speech.

Headset audio for these isolated speech segments is very clean, with almost no background noise. The distant microphone audio for these segments contains background noise in addition to speech. To remove this background noise and create clean reverberant spatial images of speech at the distant microphone, we estimated a linear time-invariant filter $\hat{h}$. This filter is estimated using least-squares to map the segmented headset speech to the noisy distant microphone observation. The filter is found using the following equation:
\begin{equation}
    \hat {h} = \arg\min_{h \in \mathbb{R}^N} \left\| x[n] * h[n] - y[n] \right\|^2,
\end{equation}
where $h[n]$ is an $N$-point causal finite impulse response filter, $x[n]$ is the headset microphone signal, $y[n]$ is the  distant microphone signal and $*$ denotes convolution. The filtered headset signal is given by $x[n] * \hat{h}[n]$. The filter was chosen to be \SI{200}{\milli\second} long ($N=3200$ at 16 kHz sampling rate), and the estimation was done in the time domain. This filtering is not perfect: if the filtered headset is subtracted from the microphone array, the residual still contains some speech content, likely due to movement and imperfect estimation. However, the filtered audio generally sounds perceptually reasonable.

Using noisy distant microphone audio and filtered headset audio, we constructed synthetic mixtures as follows. First, we generate short clips from the segments: segments shorter than \SI{5}{\second} are considered short ``complete'' clips, and segments longer than \SI{5}{\second} are chopped into non-overlapping \SI{5}{\second} clips, where any remainders of a segment less than \SI{5}{\second} are considered ``remainder'' clips. To construct an example, a \SI{5}{\second} clip is randomly chosen. The filtered headset for this clip is used as one reference source, and the residual (subtracting the filtered headset from the noisy distance microphone) is used as an imperfect noise reference that still contains traces of speech.

Next, we sample a second clip, with overlap that matches the distribution of AMI. To achieve this, we measured overlaps of two or more speakers, identified using word-level annotations, and fit a log-normal distribution on these overlaps.
The overlap $\gamma$ for each clip is sampled from this distribution.
A different random speaker is selected (with care to ensure an equal balance of speaker gender), and a clip is selected for this speaker that avoids unnatural breaks of words and matches sampled overlap $\gamma$, either by placing a complete clip with duration close to $\gamma \cdot 5$ in the middle, or using part of a clip with duration greater than  $\gamma \cdot 5$ at the beginning or end. The filtered headset for this clip is used as a second speech reference source. 

In summary, synthetic AMI examples consist of three reference sources: imperfect noise reference active for the whole clip, distant speech reference 1 (S1) active for the whole clip, and distant speech reference 2 (S2) active for the proportion $\gamma$ of the clip.

\subsection{Real Overlapping AMI}

To test our models' performance on real overlapping AMI data, we selected segments (with a minimum duration of \SI{2.5}{\second}) from the AMI Corpus test set meetings wherever the corpus annotations indicated a two-speaker overlap. We took care to balance gender of the two speakers for Edinburgh (TNO test meetings only contain male speakers, and Idiap test meetings only contain female speakers). We then annotated these segments by how much cross-talk was present in each speaker's headset audio. Segments with no or minor cross-talk were selected for evaluation. This was done to obtain an imperfect reference for the target speaker (i.e., the most prominent voice or ``foreground'' speaker). Although some segments contained minor amounts of cross-talk in the headset recordings, overall, the speaker's relative volumes sounded roughly equal in the distant microphone audio. We manually identified 92 such examples.

\subsection{MUSHRA and MUSHIRA}

The MUSHRA listening test depends on a pristine reference signal to indicate what the target signal is and to calibrate ratings (raters are required to annotate the hidden reference as 100). Because of this, MUSHRA is unsuitable to use for real recordings where we do not have pristine ground-truth. For example, MUSHRA is unsuitable to be used for real AMI recordings, because cross-talk from other speakers is present in AMI headset audio.

To solve this problem, we propose a modified version of MUSHRA which we call MUSHIRA. MUSHIRA is a slight modification of MUSHRA where raters are not required to rate the hidden reference as 100. They are instructed to rate clips according to the most prominent speech from a single speaker (the ``foreground speech''), and that the presented imperfect reference signal is one of the treatments presented. MUSHIRA is useful when a particular model can outperform the reference, e.g. if the reference is a headset with cross-talk, a perfect recovery of the foreground speech by the separation model may score higher than the reference.

For MUSHRA on synthetic AMI, we chose a subset of the full synthetic AMI set, which consists of 70 examples from each of the 3 AMI rooms. For each example, we create two items to be rated: one with speaker 1 as the reference, and one with speaker 2 as the reference. The close-talking headset is the reference, filtered headset and noisy distant mixture are anchors, and we evaluate a set of 10 models (the last 10 rows of Table \ref{tab:results}).
For MUSHIRA on real AMI, we use the headset with cross-talk as the imperfect reference, filtered headset and noisy distant mixture as anchors, and the same 10 models used for MUSHRA on synthetic AMI. For both MUSHRA and MUSHIRA, we collect 5 ratings per item.

\vspace{-5pt}
\section{Results}
\vspace{-4pt}

\begin{table*}[t]
\centering
\caption{Averaged results over synthetic and real AMI datasets. ``S1'' refers to the full-duration speaker and ``S2'' refers to the overlapping speaker. For full synthetic AMI, the absolute input SI-SNRs are \SI{0.5}{\decibel} for S1 and \SI{-9.2}{\decibel} for S2, which are used in the SI-SNRi computation. ``Warm Start'' indicates pretraining the model with MixIT on \SI{5800} hours of AudioSet (AS) data. 
    We denote the reference signal for each metric as (H) for headset or (FH) for headset filtered to the distant microphone. The 95\% confidence interval for MUSHRA is $\pm 1.1$, and for MUSHIRA is $\pm 2.0$. MUSHRA scores with a $^*$ indicate scores from a different preliminary MUSHRA study with 95\% confidence interval 1.0, with S1(H) and S2(H) scores of 64.8 and 61.6 for FH anchor, and 34.4 and 21.8 for distant mic anchor.
    }
\setlength{\tabcolsep}{4pt}
\begin{tabular}{cccccccccccccccc}
\toprule
\multicolumn{4}{c}{\bf\makecell{Model Configuration}}
& \phantom{a}
& \multicolumn{2}{c}{\bf\makecell{Full Synthetic\\AMI Dataset}}
& \phantom{a}
& \multicolumn{6}{c}{\bf\makecell{Subset Synthetic\\AMI Dataset}}
& \phantom{a}
& \multicolumn{1}{c}{\bf\makecell{Real\\AMI Dataset}}
\\
\cmidrule{1-4} \cmidrule{6-7} \cmidrule{9-14} \cmidrule{16-13} 
\makecell{Sup.\\PIT}
& \makecell{Unsup.\\MixIT}
& \makecell{Warm\\Start}
& Basis
&
& \makecell[c]{SI-SNRi \\ S1(FH)}
& \makecell[c]{SI-SNRi \\ S2(FH)}
&
& \multicolumn{2}{c}{\makecell[c]{PESQ S1\\(H)\hspace{3ex}(FH)}}
& \multicolumn{2}{c}{\makecell[c]{PESQ S2\\(H)\hspace{3ex}(FH)}}
& \multicolumn{2}{c}{\makecell[c]{MUSHRA\\S1(H)\hspace{1ex}S2(H)}}
&
& \makecell[c]{MUSHIRA}
\\
\midrule
\multicolumn{4}{c}{Headset (H)} & & -- & -- & & 4.50 & 2.18 & 4.50 & 2.56 & 100.0 & 100.0 & & 89.7 \\
\multicolumn{4}{c}{Headset filtered to distant mic (FH)} & & $\infty$ & $\infty$ & & 2.38 & 4.50 & 2.79 & 4.50 & 62.2 & 60.1 & & 46.8  \\
\multicolumn{4}{c}{Distant mic} & & 0.0 & 0.0 & & 1.69 & 1.97 & 1.62 & 1.69 & 36.0 & 25.2 & & 38.0  \\
\midrule
                            RL2M & -- & -- & learn &        &          -0.1 &            2.8 &        &           1.68 &           1.93 &           1.69 &           1.81 &           31.0$^*$ &           19.9$^*$ &        &             -- \\
                             RL2M & -- & -- & STFT &        &           1.3 &            4.3 &        &           1.65 &           1.87 &           1.65 &           1.79 &             -- &             -- &        &             -- \\
                            RL2M & -- & AS & learn &        &           1.7 &            5.7 &        &           1.84 &           2.21 &  \textbf{1.88} &           2.10 &           41.4$^*$ &           26.8$^*$ &        &             -- \\
                             RL2M & -- & AS & STFT &        &          -0.6 &            2.8 &        &           1.73 &           1.98 &           1.76 &           1.90 &             -- &             -- &        &             -- \\
                              -- & -- & AS & learn &        &           4.0 &           10.6 &        &           1.90 &           2.31 &  \textbf{1.88} &           2.06 &           47.4 &           27.4 &        &           42.5 \\
                               -- & -- & AS & STFT &        &           3.7 &            9.5 &        &           1.86 &           2.23 &           1.86 &           2.04 &           46.0 &           28.0 &        &           43.8 \\
                             -- & AMI & -- & learn &        &           3.9 &           10.6 &        &           1.80 &           2.20 &           1.81 &           2.01 &           44.5 &           27.5 &        &           40.5 \\
                              -- & AMI & -- & STFT &        &           1.0 &            5.7 &        &           1.60 &           1.82 &           1.61 &           1.73 &           34.3 &           22.4 &        &           35.5 \\
                             -- & AMI & AS & learn &        &           3.8 &            9.7 &        &           1.85 &           2.26 &           1.79 &           1.98 &           45.5 &           27.0 &        &           42.1 \\
                              -- & AMI & AS & STFT &        &           3.7 &           10.9 &        &           1.83 &           2.24 &           1.80 &           2.02 &           45.7 &           28.8 &        &           42.5 \\
                           RL2M & AMI & -- & learn &        &           4.2 &           11.4 &        &           1.86 &           2.27 &           1.84 &           2.04 &           46.0 &           29.3 &        &           41.9 \\
                            RL2M & AMI & -- & STFT &        &           2.6 &            8.2 &        &           1.78 &           2.11 &           1.76 &           1.93 &           42.5 &           26.5 &        &           41.5 \\
                           RL2M & AMI & AS & learn &        &  \textbf{4.9} &  \textbf{12.4} &        &  \textbf{1.93} &  \textbf{2.39} &  \textbf{1.88} &  \textbf{2.13} &           48.3 &           29.7 &        &           43.9 \\
                            RL2M & AMI & AS & STFT &        &           3.9 &           10.8 &        &           1.89 &           2.33 &  \textbf{1.88} &           2.12 &  \textbf{49.7} &  \textbf{29.8} &        &  \textbf{44.4} \\
\bottomrule
\end{tabular}
    \vspace{-10pt}
    \label{tab:results}
\end{table*}

In our initial experiments, we compared using the original Libri2Mix dataset \cite{cosentino2020librimix} versus our reverberant version, RLibri2Mix, for supervised training. Using the original Libri2Mix resulted in poor performance compared to RLibri2Mix in terms of SI-SNRi on our full synthetic AMI dataset: 
training with Libri2Mix yielded \SI{-2.3}{\decibel} without AudioSet warm-start and \SI{-1.8}{\decibel} with AudioSet warm-start, and RLibri2Mix yielded \SI{1.3}{\decibel} without  AudioSet warm-start and \SI{1.7}{\decibel} with AudioSet warm-start. Thus, we only use RLibri2Mix as supervised data in our main experiments.

Despite matching the AMI data in terms of the presence of reverb, RLibri2Mix is mismatched in other ways.  RLibri2Mix has synthetic reverberation rather than real, it consists of read rather than spontaneous speech, it contains no background noise, and the pattern of overlap between speakers is different. Our goal with using RLibri2Mix was to use a relatively standard supervised dataset and demonstrate that models trained with mismatched data can be adapted using unsupervised learning. With more engineering effort and knowledge of the task domain, better-matching synthetic datasets could be constructed. However the goal here is to simulate more realistic conditions where such knowledge is not available.

We also performed an initial evaluation with 
a MUSHRA listening test on a different, preliminary version of the synthetic AMI dataset. This evaluation only included models using learnable basis, with purely-supervised training on RLibri2Mix, with or without an AudioSet warm start, as well as models using unsupervised and semi-supervised training. The purely-supervised models were the lowest scoring models in terms of MUSHRA for this evaluation: 
\SI{25.5} without AudioSet warm-start, \SI{34.1} with AudioSet warm-start, compared to \SI{28.1} for distant microphone anchor, \SI{35.6} for the next best model (unsupervised MixIT on AMI without AudioSet warm-start), and \SI{63.2} for the filtered headset anchor. Thus we decided to exclude them in our final MUSHRA and MUSHIRA evaluations so that we could directly compare STFT and learnable basis.

Table \ref{tab:results} shows our overall results in terms of SI-SNRi on the full synthetic AMI dataset (using filtered headset as reference, which is analogous to a {\tt bss\_eval}-like filtering of the reference signal \cite{vincent2006performance}), PESQ and MUSHRA scores evaluated on the subset of synthetic AMI examples used for the MUSHRA listening test, and MUSHIRA scores for real overlapping AMI examples. We measure PESQ using either the headset (H) or filtered headset (FH) as the reference signal, to match MUSHRA and SI-SNRi computations, respectively.

Note that SI-SNRi is lower for S1 compared to S2. This is because S1 has a higher average input SI-SNR of 0.5 dB, versus -9.2 dB for S2. Also, the S1 filtered-headset reference is imperfect, since the time-invariant filtering projection from headset to distant microphone is not perfect. The perceptual PESQ and listening test scores indicate that models are able to separate these filtered-headset references relatively well, although no method exceeds the performance of the filtered headset anchor.

Overall, using a warm-start from pretraining on AudioSet provides a consistent improvement across training configurations in terms of all metrics. These pretrained models exhibit surprisingly strong performance even without tuning on additional datasets. Training on AMI alone with MixIT only slightly underperforms training on open-domain AudioSet data, which is remarkable given that the AMI data is about two orders of magnitude smaller (tens of hours, versus 5800 hours of AudioSet data), highlighting the benefit of in-domain training. In future work, we intend to study the effect of relative dataset sizes on separation performance.

Generally, using a learnable basis yields higher SI-SNRi compared to STFT, but learnable basis models produced slightly lower MUSHRA and MUSHIRA scores.  Informal listening identified more artifacts in learnable basis models as a possible cause.

The best training configuration is semi-supervised (PIT on reverberant Libri2Mix, MixIT on AMI), warm-starting from weights pretrained with MixIT on AudioSet. We think that these training methods are complementary: even though it is mismatched, supervised data provides examples of isolated speech that are lacking in the unsupervised AMI MoMs, while the matched AMI MoMs help adapt the model towards the target domain. 

Audio demos are available online at  \url{https://ami-mixit.github.io}.

\section{Conclusion}
\label{sec:conclusion}

In this paper, we used MixIT to train separation models targeted towards real-world meeting data. Our best results used pretraining with MixIT on a large amount of open-domain data from AudioSet \cite{AudioSet}, followed by fine-tuning with PIT on supervised data (a reverberant version of Libri2Mix \cite{cosentino2020librimix}) and MixIT on unsupervised data (real distant microphone recordings from AMI \cite{CarlettaJ2006AMI}). To estimate objective performance, we constructed a synthetic version of AMI that takes advantage of its annotations and parallel headset and distant microphone recordings. We also proposed a generalization of MUSHRA called MUSHIRA to facilitate human evaluation of source separation systems with imperfect reference signals.

We hope to extend this work by further investigating dereverberation, as well as taking advantage of multiple microphones (the AMI data has a 8-microphone circular array that is consistent across locations). Fine-tuning our models trained with MixIT on AudioSet on other downstream tasks is another interesting avenue of future work.

\noindent
{\bf Acknowledgements:} Thanks to Kevin Wilson for helpful comments on the manuscript.

\vfill\pagebreak

\bibliographystyle{IEEEbib_nourl}
\balance
\bibliography{refs}

\begin{thebibliography}{10}
\def\url#1{}
\csname url@samestyle\endcsname
\providecommand{\newblock}{\relax}
\providecommand{\bibinfo}[2]{#2}
\providecommand{\BIBentrySTDinterwordspacing}{\spaceskip=0pt\relax}
\providecommand{\BIBentryALTinterwordstretchfactor}{4}
\providecommand{\BIBentryALTinterwordspacing}{\spaceskip=\fontdimen2\font plus
\BIBentryALTinterwordstretchfactor\fontdimen3\font minus
  \fontdimen4\font\relax}
\providecommand{\BIBforeignlanguage}[2]{{%
\expandafter\ifx\csname l@#1\endcsname\relax
\typeout{** WARNING: IEEEtran.bst: No hyphenation pattern has been}%
\typeout{** loaded for the language `#1'. Using the pattern for}%
\typeout{** the default language instead.}%
\else
\language=\csname l@#1\endcsname
\fi
#2}}
\providecommand{\BIBdecl}{\relax}
\BIBdecl

\bibitem{LuoY2019convtasnet}
Y.~Luo and N.~Mesgarani, ``{Conv-TasNet}: Surpassing ideal time--frequency
  magnitude masking for speech separation,'' \emph{IEEE/ACM Transactions on
  Audio, Speech, and Language Processing}, vol.~27, no.~8, 2019.

\bibitem{Chen2020dual}
J.~Chen, Q.~Mao, and D.~Liu, ``{Dual-Path Transformer Network: Direct
  Context-Aware Modeling for End-to-End Monaural Speech Separation},'' in
  \emph{Proc. Interspeech}, 2020.

\bibitem{HuangP2015ieeeacmaslp}
P.-S. Huang, M.~Kim, M.~Hasegawa-Johnson, and P.~Smaragdis, ``Joint
  optimization of masks and deep recurrent neural networks for monaural source
  separation,'' \emph{IEEE/ACM Transactions on Audio, Speech, and Language
  Processing}, vol.~23, no.~12, Dec 2015.

\bibitem{hershey2016deep}
J.~R. Hershey, Z.~Chen, J.~Le~Roux, and S.~Watanabe, ``{Deep clustering:
  Discriminative embeddings for segmentation and separation},'' in \emph{Proc.
  {IEEE} International Conference on Acoustics, Speech, and Signal Processing
  ({ICASSP})}, 2016.

\bibitem{yu2017permutation}
D.~Yu, M.~Kolb{\ae}k, Z.-H. Tan, and J.~Jensen, ``{Permutation invariant
  training of deep models for speaker-independent multi-talker speech
  separation},'' in \emph{Proc. {IEEE} International Conference on Acoustics,
  Speech, and Signal Processing ({ICASSP})}, 2017.

\bibitem{wang2018supervised}
D.~L. Wang and J.~Chen, ``{Supervised Speech Separation Based on Deep Learning:
  An Overview},'' \emph{{IEEE/ACM} Transactions on Audio, Speech, and Language
  Processing}, vol.~26, no.~10, 2018.

\bibitem{KinoshitaK2013REVERB}
K.~Kinoshita, M.~Delcroix, T.~Yoshioka, T.~Nakatani, E.~Habets, R.~Haeb-Umbach,
  V.~Leutnant, A.~Sehr, W.~Kellermann, R.~Maas \emph{et~al.}, ``{The REVERB
  challenge: A common evaluation framework for dereverberation and recognition
  of reverberant speech},'' in \emph{Proc. IEEE Workshop on Applications of
  Signal Processing to Audio and Acoustics (WASPAA)}.\hskip 1em plus 0.5em
  minus 0.4em\relax IEEE, 2013.

\bibitem{BarkerJ2015CHIME}
J.~Barker, R.~Marxer, E.~Vincent, and S.~Watanabe, ``{The third ‘CHiME’
  speech separation and recognition challenge: Dataset, task and baselines},''
  in \emph{Proc. IEEE Automatic Speech Recognition and Understanding
  Workshop}.\hskip 1em plus 0.5em minus 0.4em\relax IEEE, 2015.

\bibitem{maciejewski2018building}
M.~Maciejewski, G.~Sell, L.~P. Garcia-Perera, S.~Watanabe, and S.~Khudanpur,
  ``{Building Corpora for Single-Channel Speech Separation Across Multiple
  Domains},'' \emph{arXiv preprint arXiv:1811.02641}, 2018.

\bibitem{manilow2019cutting}
E.~Manilow, G.~Wichern, P.~Seetharaman, and J.~Le~Roux, ``{Cutting music source
  separation some Slakh: A dataset to study the impact of training data quality
  and quantity},'' in \emph{IEEE Workshop on Applications of Signal Processing
  to Audio and Acoustics (WASPAA)}, 2019.

\bibitem{SivaramanA2021interspeech}
A.~Sivaraman, S.~Kim, and M.~Kim, ``{Personalized Speech Enhancement through
  Self-Supervised Data Augmentation and Purification},'' in \emph{Proc.
  Interspeech}, 2021.

\bibitem{neri2021unsupervised}
J.~Neri, R.~Badeau, and P.~Depalle, ``Unsupervised blind source separation with
  variational auto-encoders,'' in \emph{Proc. European Signal Processing
  Conference (EUSIPCO)}, 2021.

\bibitem{wisdom2020mixit}
S.~Wisdom, E.~Tzinis, H.~Erdogan, R.~J. Weiss, K.~Wilson, and J.~R. Hershey,
  ``Unsupervised sound separation using mixture invariant training,'' in
  \emph{Advances in Neural Information Processing Systems}, 2020.

\bibitem{CarlettaJ2006AMI}
J.~Carletta, S.~Ashby, S.~Bourban, M.~Flynn, M.~Guillemot, T.~Hain, J.~Kadlec,
  V.~Karaiskos, W.~Kraaij, M.~Kronenthal, G.~Lathoud, M.~Lincoln, A.~Lisowska,
  I.~McCowan, W.~Post, D.~Reidsma, and P.~Wellner, ``{The AMI Meeting Corpus: A
  Pre-announcement},'' in \emph{{Machine Learning for Multimodal Interaction}},
  2006.

\bibitem{AudioSet}
J.~F. Gemmeke, D.~P.~W. Ellis, D.~Freedman, A.~Jansen, W.~Lawrence, R.~C.
  Moore, M.~Plakal, and M.~Ritter, ``{Audio Set: An ontology and human-labeled
  dataset for audio events},'' in \emph{Proc. {IEEE} International Conference
  on Acoustics, Speech, and Signal Processing ({ICASSP})}, 2017.

\bibitem{itu2014mushra}
ITU, ``{Method for the subjective assessment of intermediate quality level of
  audio systems},'' \emph{BS.1534}, 2014.

\bibitem{kavalerov2019universal}
I.~Kavalerov, S.~Wisdom, H.~Erdogan, B.~Patton, K.~Wilson, J.~{Le Roux}, and
  J.~R. Hershey, ``{Universal Sound Separation},'' in \emph{Proc. {IEEE}
  Workshop on Applications of Signal Processing to Audio and Acoustics
  ({WASPAA})}.\hskip 1em plus 0.5em minus 0.4em\relax IEEE, 2019.

\bibitem{kingma2014adam}
D.~P. Kingma and J.~Ba, ``{Adam: A method for stochastic optimization},'' in
  \emph{Proc. International Conference on Learning Representations ({ICLR})},
  2015.

\bibitem{cosentino2020librimix}
J.~Cosentino, M.~Pariente, S.~Cornell, A.~Deleforge, and E.~Vincent,
  ``{LibriMix}: An open-source dataset for generalizable speech separation,''
  \emph{arXiv preprint arXiv:2005.11262}, 2020.

\bibitem{wang2021sequential}
Z.-Q. Wang, H.~Erdogan, S.~Wisdom, K.~Wilson, and J.~R. Hershey, ``{Sequential
  Multi-Frame Neural Beamforming for Speech Separation and Enhancement},'' in
  \emph{Proc. Spoken Language Technology Workshop {(SLT)}}, 2021.

\bibitem{wisdom2021sparse}
S.~Wisdom, A.~Jansen, R.~J. Weiss, H.~Erdogan, and J.~R. Hershey, ``{Sparse,
  Efficient, and Semantic Mixture Invariant Training: Taming In-the-Wild
  Unsupervised Sound Separation},'' in \emph{Proc. IEEE Workshop on
  Applications of Signal Processing to Audio and Acoustics (WASPAA)}, 2021.

\bibitem{vincent2006performance}
E.~Vincent, R.~Gribonval, and C.~F{\'e}votte, ``{Performance measurement in
  blind audio source separation},'' \emph{{IEEE/ACM} Transactions on Audio,
  Speech, and Language Processing}, vol.~14, no.~4, 2006.

\end{thebibliography}

\end{document}